\documentclass[aps,pra,twocolumn,nofootinbib,superscriptaddress,floatfix]{revtex4-1}
\usepackage{epsfig}
\usepackage[colorlinks,linkcolor=blue,anchorcolor=blue,citecolor=blue,urlcolor=blue,breaklinks=true]{hyperref}
\usepackage{graphics}
\usepackage{color}
\usepackage{amsmath}
\usepackage{bm}
\usepackage{subfig}
\usepackage{braket}             
\usepackage{tikz}
\usepackage[margin=1in]{geometry}

\begin{document}
\author{Emil A. Gazazyan}
\affiliation{Institute for Physical Research, of the National Academy of Sciences of the Republic of Armenia, Ashtarak-2, 0203, Republic of Armenia}
\affiliation{Institute for Informatics and Automation Problems, of the National Academy of Sciences of the Republic of Armenia, Yerevan, 0014, 1, P. Sevak str., Republic of Armenia}
\author{Gayane G. Grigoryan}
\email{gaygrig@gmail.com}
\affiliation{Institute for Physical Research, of the National Academy of Sciences of the Republic of Armenia, Ashtarak-2, 0203, Republic of Armenia}
\author{Vanush Paturyan}
\affiliation{Department of Computer Science, Maynooth University, Maynooth, Co Kildare, Ireland}

\title{Stability of Adiabatic States in a Dissipative Three-Level System}

\begin{abstract}
	The necessary and sufficient conditions for the stability of adiabatic states in three-level quantum systems are investigated
	analytically and numerically. Various possible configurations of three-level systems under exact two-photon resonance are considered.
	It is shown that in all these schemes, the lifetime of the studied states is determined by the dephasing time between levels that are not connected by a dipole transition.
	An efficient population transfer through the b-state at room temperature is demonstrated despite relatively long relaxation times. It is also demonstrated that, in case of large one-photon detuning, the so-called b-state has the same lifetime as that of the dark state.
	The evolution of adiabatic states for arbitrary values of single-photon detunings has been studied numerically.
\end{abstract}

\pacs{}
\maketitle

\section{Introduction}
Coherent manipulation and control of quantum systems interacting with laser radiation form the basis for quantum technologies and various
quantum applications (see \cite{VanFrank2016,Koch2022,Ansel2024}  and references therein). In this context, resonant interactions
\cite{Safronova2018,Vitanov2017}  have attracted significant attention in recent years
\cite{YeZoller2024,Bouteyre2025,Breu2025,Bluvstein2024,Bjorkman2025,Sargsyan2019,Wadenpfuhl2025}.Among them, adiabatic interactions are of
particular importance due to their robust and versatile applicability, especially in the field of quantum information processing \cite{13,14,15,16}.

In closed quantum systems interacting adiabatically with resonant radiation the eigenstates of the interaction Hamiltonian are formed, giving rise to a wide range of coherent phenomena that have been extensively explored in the literature \cite{Cooper2019,Shore2017,Shore1990,TerMikaelyan1997}.

The lifetime of such adiabatic states in closed systems is governed by the inequalities
${1} / {| \lambda_i - \lambda_j |}_{max} \ll T$, where $\lambda{i,j}$ are the eigenvalues of the interaction hamiltonian (or the quasienergies if the system). It is worth recalling that the quasienergies $\lambda_k$ represent the energies of the
“atom + field" system, and the eigenstates $|\lambda_k \rangle$ in turn describe the states of the "atom + field" system at a particular energy level.
Conventionally, the labeling of these states is chosen such that they assymoptotically reduce to the bare atomic ground states in the limit of the vanishing field amplitudes.

In contrast, in open quantum systems where various relaxation and decoherence processes are present adiabatic states that were formed in the system can decay over time.

\begin{figure}[ht]
	\centering
	\begin{tikzpicture}[scale=1.1]

		\begin{scope}
		\draw[thick] (0,0) -- (1,0) node[below] {1};
		\draw[thick] (2,0) -- (3,0) node[below] {2};
		\draw[thick] (1,2.5) -- (2,2.5) node[above] {3};

		\draw[->, thick] (0.5,0) -- (1.5,2.3) node[pos=0.5, xshift=8pt] {\(\Omega_p\)};
		\draw[->, thick] (2.3,0) -- (1.5,2.3) node[pos=0.5, xshift=-8pt] {\(\Omega_c\)};

		\draw[->, dashed, thick] (1.5,2.5) -- (0.2,0) node[midway, left] {\(\gamma_1\)};
		\draw[->, dashed, thick] (1.5,2.5) -- (2.7,0) node[midway, right] {\(\gamma_2\)};

		\draw[dashed, blue] (0.9,2.3) -- (2.1,2.3);
		\node at (2.3,2.3) {\(\Delta\)};

		\node at (1.5,-0.8) {$\Lambda$-system};
        \end{scope}
		\begin{scope}[xshift=4cm]
			\draw[thick] (0,0) -- (1,0) node[below] {1};
			\draw[thick] (0,1.5) -- (1,1.5) node[right] {3};
			\draw[thick] (0,3) -- (1,3) node[above] {2};

			\draw[->, thick] (0.5,0) -- (0.5,1.3) node[midway, right] {\(\Omega_p\)};
			\draw[->, thick] (0.5,1.3) -- (0.5,3) node[midway, right] {\(\Omega_c\)};

			\draw[->, dashed, thick] (0.3,3) -- (0.3,1.5) node[midway, left] {\(\gamma_2\)};
			\draw[->, dashed, thick] (0.3,1.5) -- (0.3,0) node[midway, left] {\(\gamma_1\)};

			\draw[dashed, blue] (-0.2,1.3) -- (1.2,1.3);
			\node at (1.4,1.1) {\(\Delta\)};

			\node at (0.5,-0.8) {$\Xi$-system};
		\end{scope}

		\begin{scope}[xshift=2cm, yshift=-4cm]
			\draw[thick] (0,2.5) -- (1,2.5) node[above] {1};
			\draw[thick] (2,2.5) -- (3,2.5) node[above] {2};
			\draw[thick] (1,0) -- (2,0) node[below] {3};

			\draw[->, thick] (1.5,0) -- (0.5,2.2) node[pos=0.5, xshift=8pt] {\(\Omega_p\)};
			\draw[->, thick] (1.5,0) -- (2.5,2.2) node[pos=0.5, xshift=-7pt] {\(\Omega_c\)};

			\draw[->, dashed, thick] (2.5,2.5) -- (2,0) node[midway, right] {\(\gamma_2\)};
			\draw[->, dashed, thick] (0.5,2.5) -- (1,0) node[midway, left] {\(\gamma_1\)};

			\draw[dashed, green!70!black] (0.2,2.3) -- (2.9,2.3);
			\node at (3.1,2.2) {\(\Delta\)};

			\node at (1.5,-0.8) {$V$-system};
		\end{scope}

	\end{tikzpicture}
	\caption{Three-level quantum system configurations: $\Lambda$, $\Xi$, and $V$ schemes with respective couplings and decay channels.
	The unusual numbering of levels was chosen to enable the analysis of all three systems within the same mathematical formalism.
	}
	\label{Fig:MultilevelSystem}
\end{figure}
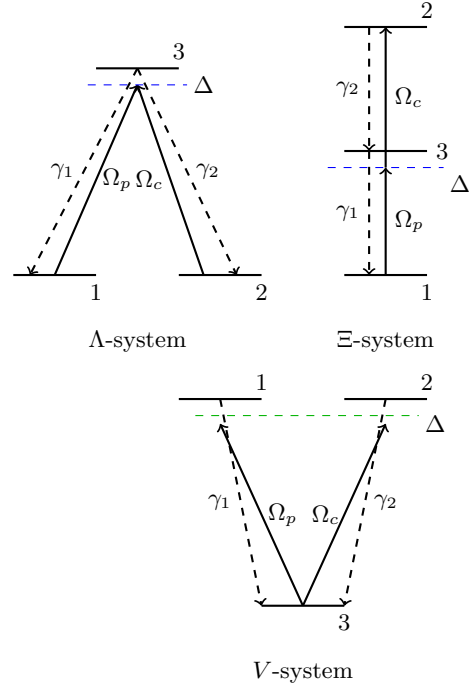

The primary goal of the present work is to determine the characteristic timescales over which performance or the quantum logic gates based on the
adiabatic interactions (such as the CNOT gate) may be regarded as efficient. To this end, both analytical and numerical analyses
are conducted on adiabatic states in all three configurations of three-level quantum systems (see Fig. \ref{Fig:MultilevelSystem}).
The inequalities that determine the regions of variation of interaction parameters so that the adiabatic states remain stablehave been obtained.
Furthermore, a detailed analysis of the population transfer within the system is presented, illustrating the
practical implications of adiabatic control in the presence
of relaxation.

\section{Mathematical Formalism}

In the basis of eigenstates  $|\lambda_i\rangle$ , the interaction Hamiltonian is diagonal
$	H = \sum_i \lambda_i |\lambda_i\rangle \langle \lambda_i|$. Consequently, in the basis of atomic states $\{e_i\}$,
the Hamiltonian $H' = U^{\dagger} H U, \qquad H' = \sum_i \lambda_i \, |e_i\rangle \langle e_i| $ becomes diagonal.
Here $U$ is the transformation matrix from the atomic basis $\left\{ |e_k\rangle\right\}$ to
the basis of the Hamiltonian's eigenvalues $\left\{ | \lambda_k\rangle \right \}$ (i.e. $U|e_k\rangle = |\lambda_k\rangle$),
and the matrix elements are given by $U_{ik} = \langle e_i \,|\, \lambda_k \rangle$
that is, the columns of the matrix $U$ represent the eigenvectors $|\lambda_k\rangle$.

In this basis the equation of the density matrix becomes:

\begin{equation}
	\frac{dR}{dt} = -i[H', R] + L' + [R, F]
\label{eq:dRdt}
\end{equation}

Here $L^{'} = U^{\dagger} L U$, where $L$ is the Lindblad operator \cite{Nathan2020,22} associated with various relaxation and dephasing processes.
The matrix $F = U^{\dagger} \frac{dU}{dt}$ characterizes nonadiabatic transitions and is an antisymmetric matrix,
with $F_{jl} = -\langle\lambda_j|\dot{\lambda_l}\rangle, F_{ii} = 0$, $F_{ij} = -F_{ji}$.
If the quasienergies of the system do not depend on time, then $F \equiv 0$.
In this case, for short interaction times, when relaxation processes can be neglected (i.e., $L \equiv 0$),
the system of equations (\ref{eq:dRdt}) has a simple solution:

\begin{equation}
	R_{ij} = R_{ij}(0)\,\exp\!\big(-i(\lambda_i - \lambda_j)t\big).
\label{eq:Rsol}
\end{equation}

With the initial conditions $R_{nn}(0) = 1$, $R_{ij}(0) = 0$ for $i,j \neq n$ (corresponding to the condition
that at the initial moment one of the adiabatic states is formed), solution (\ref{eq:Rsol}) becomes
$R = \sigma_{nn} = |e_n\rangle\langle e_n|$.

In the general case, the system of equations (\ref{eq:dRdt}) is rather cumbersome; however,
we are only interested in solutions of the form $R = \sigma_{nn} + \varepsilon R^{'}$,
where $\varepsilon$ is a small dimensionless parameter to be determined in the course of solving the problem.

We also provide another relation that will be useful later: under the condition $R = \sigma_{nn}$,
the density matrix is $\rho = |\lambda_n\rangle\langle \lambda_n|$.

\section{Three-Level System}

For a three-level  system in exact two-photon resonance (i.e. $\omega_{21} = \omega_1 - \omega_2 + \dot{\varphi}_2 - \dot{\varphi}_1$),
the eigenvalues are given by:$ \lambda_1 = 0, \quad \lambda_{2,3} = \frac{1}{2} \left( \Delta \mp \sqrt{\Delta^2 + 4\Omega^2} \right)$
We adopt  standard notations from the literature:
$
	\tan\theta = \frac{\Omega_p}{\Omega_c}, \quad \tan2\Phi = \frac{2\Omega}{\Delta}, \quad \Omega = \sqrt{\Omega_c^2 + \Omega_p^2}
$
Then the eigenstates of the interaction Hamiltonian are:

\begin{align*}
		\ket{\lambda_1} & = \cos\theta \ket{1} - \sin\theta \ket{2}                                      \\
		\ket{\lambda_2} & = \sin\theta \cos\Phi \ket{1} + \cos\theta \cos\Phi \ket{2} - \sin\Phi \ket{3} \\
		\ket{\lambda_3} & = \sin\theta \sin\Phi \ket{1} + \cos\theta \sin\Phi \ket{2} + \cos\Phi \ket{3}
\end{align*}

We begin by analyzing the well-known “dark state” $\ket{\lambda_1}$.
The Lindblad operator for this system is known (see e.g., \cite{23}) and given in the appendix. In the new basis,
this operator looks similar:

\begin{equation}
\begin{split}
		L' & = \frac{\gamma_{1}}{2}(2S_{13}RS_{31} - S_{33}R - RS_{33}) + \\
		   &   \frac{\gamma_{2}}{2}(2S_{23}RS_{32} - S_{33}R - RS_{33}) + \\
		   &   \frac{\gamma_{2\text{deph}}}{2}(2S_{22}RS_{22} - S_{22}R - RS_{22}) +\\
		   &   \frac{\gamma_{3\text{deph}}}{2}(2S_{33}RS_{33} - S_{33}R - RS_{33})
\end{split}
\label{eq:lindblad}
\end{equation}

Here, $S_{ij} = U^{\dagger} \sigma_{ij} U$, and $\gamma_{1}, \gamma_{2}$ are spontaneous emission rates at the corresponding transitions,
while $\gamma_{2\text{deph}}, \gamma_{3\text{deph}}$ are dephasing rates at the same transitions(see Fig. \ref{Fig:MultilevelSystem}).

From equation (\ref{eq:dRdt}), we obtain:
\begin{equation}
 \begin{split}
	R_{11} & = 1 - \frac{\gamma_c}{2} \int_0^t \sin^2 2\theta \, dt' \\
	R_{22} & = \frac{\gamma_c}{2} \int_0^t \sin^2 2\theta \cos^2\Phi \, dt'
 \end{split}
\label{eq:rdiag}
\end{equation}

\begin{equation}
 \begin{split}
	R_{21} & = \int_0^t \left( \frac{\gamma_c}{4} \sin 4\theta - \dot{\theta} \right) \cos\Phi \, e^{i\int_{t'}^t (\lambda_2 - \lambda_1) dt''} \, dt' \\
	R_{31} & = \int_0^t \left( \frac{\gamma_c}{4} \sin 4\theta - \dot{\theta} \right) \sin\Phi \, e^{i\int_{t'}^t (\lambda_3 - \lambda_1) dt''} \, dt' \\
	R_{32} & = \int_0^t \left( \frac{\gamma_c}{4} \sin^2 2\theta \sin 2\Phi \right) \, e^{i\int_{t'}^t (\lambda_3 - \lambda_2) dt''} \, dt'
  \end{split}
 \label{eq:rnondiag}
\end{equation}

As expected, the state $\ket{\lambda_1}$ is insensitive to relaxation at the upper level and depends only on the
lower state coherence decay rate $\gamma_c$. A necessary (but not sufficient) condition for stability is $\gamma_c T \ll 1$, where $T$ is the interaction time.

Regarding equations \eqref{eq:rnondiag}, they can be minimized due to the rapidly
oscillating exponential under the integral (see, for example, \cite{24,25}).
As is well known, the integral of a rapidly oscillating function $	\int a e^{i\beta t}\, dt \notag $
can be represented as a series expansion in powers of $\beta^{-1}$.
Finally, the necessary and sufficient conditions for the stability of the
state $\lvert \lambda_1 \rangle$ are given by

\begin{equation}
	\gamma_c T \ll 1,
	\quad T\sqrt{\Delta^{2} + 4\Omega^{2}} \gg 1,
	\quad \frac{\Omega^{2} T}{\Delta} \gg 1.
	\label{eq:stcnd}
\end{equation}

The last two conditions are the conditions of adiabaticity of the interaction,
while the first condition is a limitation on the system's decoherence time.
In some works (see, for example, \cite{26,27}), it is proposed to use pulses
satisfying the condition $ \sin 4\theta - \dot{\theta} = 0, \notag \quad \text{i.e., } \tan2\theta \sim \tan2\theta_{0}\,\notag
	e^{\gamma_c (t - t_{0})} \notag$.
Such pulses indeed lead to medium transparency for the $3 \to 1$ transition.
However, under the condition $\gamma_{c}T \ll 1$, which is necessary for the
stability of the first adiabatic state, the use of such pulses loses its meaning.

Note that under conditions (\ref{eq:stcnd})
the matrix elements of the transitions $\rho_{31}$ and $\rho_{32}$ vanish.
Thus, in a medium where this state is realized, the pulses propagate without
absorption and without distortion of their shape.

Let us now consider how conditions (\ref{eq:stcnd}) are modified in the case of realizing
the second adiabatic state $\lvert \lambda_{2} \rangle$.
Restricting ourselves to the case of large one-photon detunings and neglecting
terms proportional to $\Phi^{2}$, from system (\ref{eq:dRdt}) we obtain the following solutions:

\begin{equation}
 \begin{split}
	R_{11} &= \frac{\gamma_{c}}{2} \int_{0}^{t} \sin^{2}2\theta \, dt', \\
	R_{22} &= 1 - \frac{\gamma_{c}}{2} \int_{0}^{t} \sin^{2}2\theta \, dt',
 \end{split}
 \label{eq:Rsolad}
\end{equation}

Thus, as in the previous case, the condition $\gamma_{c}T \ll 1$ is necessary
for the stability of the second adiabatic state. We emphasize that in the expressions for the
populations of quasi-energy states, the fast relaxations from the third level do not appear, although the state
$\lvert \lambda_{2} \rangle$ contains it.

The off-diagonal elements $R_{12}$ and $R_{31}$, which describe transitions
between the corresponding adiabatic states, also do not depend on fast relaxations:

\begin{equation}
 \begin{split}
	R_{12} = \int_0^t \left(\frac{\gamma_c}{4}\sin 4\theta - \dot{\theta}\right)
	e^{\int_{t}^{t'} i(\lambda_{2}-\lambda_{1})dt''} \, dt', \\
	R_{31} = \int_0^t \gamma_{c}\,\Phi \sin^{2}\theta \cos^{2}\theta
	e^{\int_{t}^{t'} i(\lambda_{3}-\lambda_{1})dt''} \, dt'.\\
  \end{split}
 \label{eq:Rsolnondiag}
\end{equation}

Taking into account that $\gamma_{c}\Phi$ is a quantity of second-order smallness,
we may assume that $R_{31} \approx 0$.
However, for the $3 \to 2$ transition, dependence on fast relaxations appears:

\begin{equation}
	R_{32} = \int_0^t \left(\Gamma_{1}\Phi + \gamma_{c}\Phi \cos^{4}\theta + \dot{\Phi}\right)
	e^{\int_{t}^{t''} i(\lambda_{3}-\lambda_{2})dt''} \, dt'.
	\label{eq:R32}
\end{equation}

Given that $\gamma_{c} \ll \Gamma_{1}$, the prefactor in the integrand can be
approximated as $\Gamma_{1}\Phi + \dot{\Phi}$.
Setting this expression to zero yields the law of temporal variation of the angle $\Phi$
under which $R_{32} \equiv 0$, namely $	\Phi(t) = C \, e^{-\Gamma_1 t}.$
Such a regime may be realized, for example, by changing the detuning according to
$\Delta(t) = \Delta_{0} \Omega(t) e^{\Gamma_1(t-t_{0})},$
where $t_{0}$ is a certain moment of time after switching on the interaction.

Although such a choice of a varying detuning ensures high accuracy, in experiment
it may be difficult to implement. A simpler requirement is the condition
$	\Gamma_{1} \ll (\lambda_{3}-\lambda_{2}) \approx \Delta,$
with $\Gamma_{2} = \Gamma_{1} + \gamma_{c} \approx \Gamma_{1}$ since $\gamma_{c} \ll \Gamma_{1}$.
Thus, the stability conditions of the second adiabatic state can be written as
\begin{equation}
 \begin{split}
	\gamma_{c}T \ll 1, \\
	\quad T\sqrt{\Delta^{2}+4\Omega^{2}} \gg 1, \\
	\quad \frac{\Omega^{2}T}{\Delta} \gg 1, \\
	\quad \Gamma_{1} \ll \Delta,
 \end{split}
 \label{eq:secondstb}
\end{equation}
Under these conditions, the matrix elements $\rho_{31}$ and $\rho_{32}$ are not zero
but are real quantities. This means that the pulses propagate in the medium without
absorption, but with a certain phase velocity.

The most unstable state is the third adiabatic state $\lvert \lambda_{3} \rangle$.
\begin{equation}
	R_{33} = 1 - \gamma t
\label{eq:R33}
\end{equation}
Here, $\gamma$ is the total decay rate from the upper level.
A necessary condition is $\gamma T \ll 1$ for the $\Lambda$ system.
Recall that the total decay rate of the upper level $\gamma$ is typically
three orders of magnitude greater than $\gamma_{c}$.
The state $\lvert \lambda_{3} \rangle$ can be realized in the $\Lambda$ system
only if the atom is initially in the upper excited state before the interaction,
so this state has no practical application.

\begin{figure}[t]
	\centering
	\includegraphics[width=0.4\textwidth]{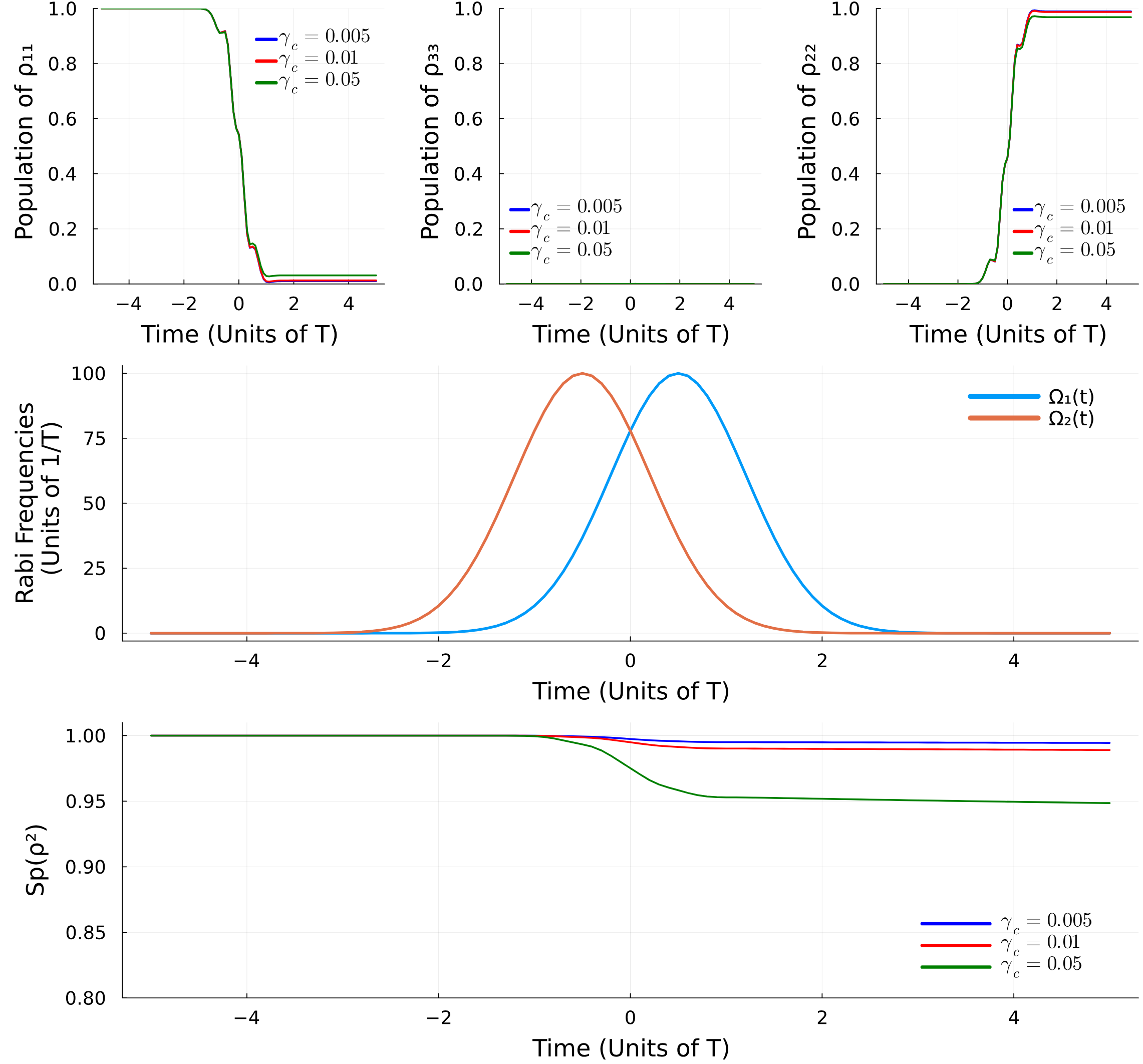}
	\caption{Numerical simulation of population transfer in a three-level system using the STIRAP.
	Parameters: spontaneous decay rates $\Gamma_1T = \Gamma_2T= 0.5$, peak Rabi frequencies $\Omega_{0,1}T=\Omega_{0,2}T = 100$,
	detunings $\Delta_1T=\Delta_2T =1000$.
	The plots show the time evolution of populations $\rho_{11}$, $\rho_{22}$, and $\rho_{33}$,
	the applied Gaussian pump and Stokes fields $\Omega_{1}(t)$ and $\Omega_{2}(t)$, and the purity of state
	$\mathrm{Sp}(\rho^2)$ for different decoherence rates $\gamma_{c}T = 0.005, 0.01,0.05$.}
\label{Fig:stirap}
\end{figure}

\begin{figure}[t]
	\centering
	\includegraphics[width=0.4\textwidth]{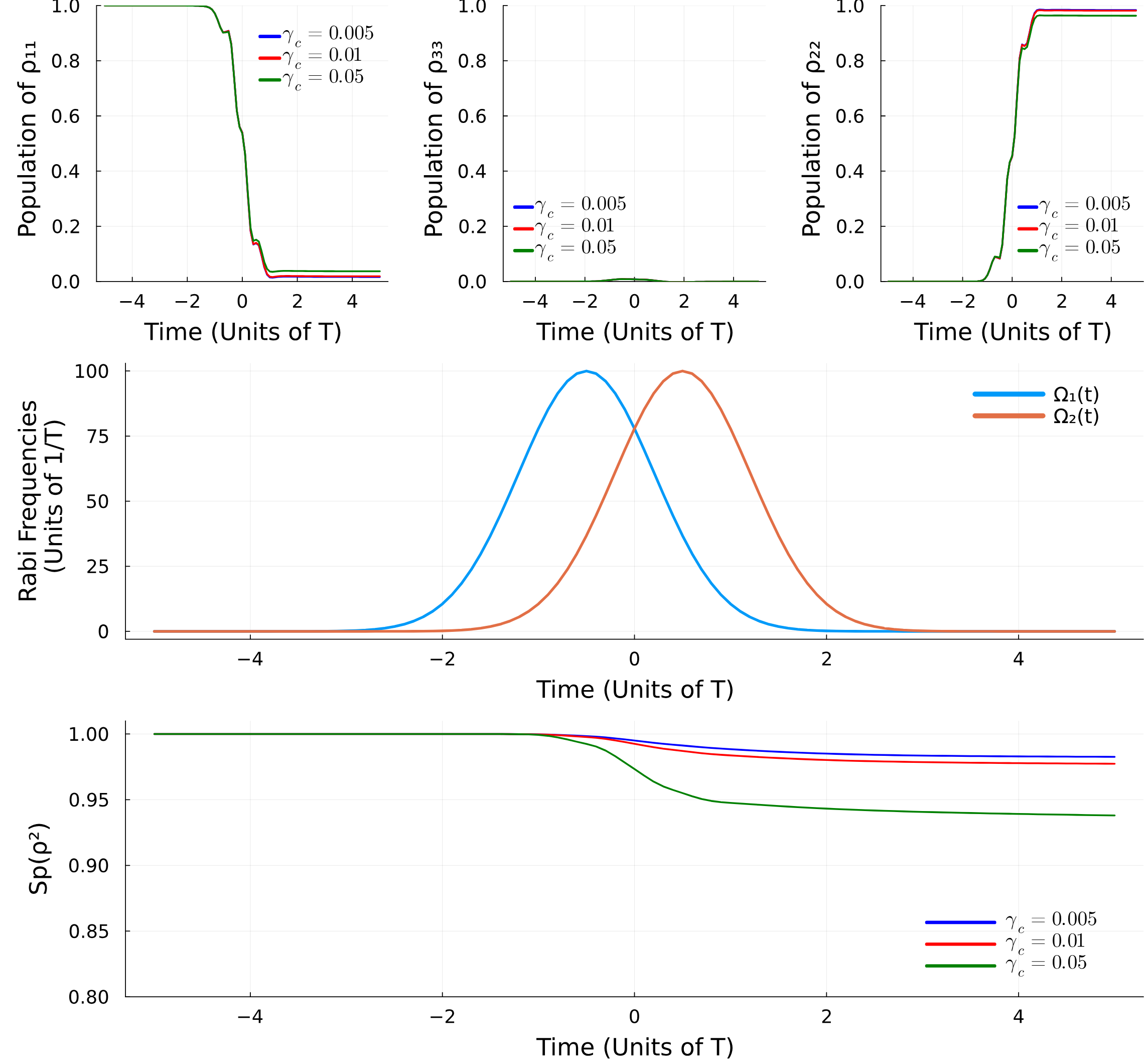}
	\caption{Numerical simulation of population transfer in a three-level system using the b-STIRAP.
	Parameters: same as Figure \ref{Fig:stirap}}
\label{Fig:bstirap}
\end{figure}

Fig. \ref{Fig:stirap} shows the population transfer by the counterintuitive pulse sequence, i.e. by Stimulated Raman Adiabatic Passage (STIRAP) method, while Fig. \ref{Fig:bstirap} shows the transfer by the intutivie pulse sequence (b-STIRAP).

Fig. \ref{Fig:sphur_delta_bstirap} shows the results of the numerical modelling of the purity of the $| \lambda_2 \rangle$ state as a dependency of the single-photon detuning. It is shown that when the detuning is decreased the mix-in of the other two states grows rapidly.

Analogous formulas can be obtained for the $\Sigma$-system and the $V$-system,
taking into account the formulas given in the Appendix.
The most significant difference in these systems is the fact that
$\gamma_{c}$ in both cases is no longer the pure dephasing rate but includes
fast spontaneous relaxations (see Appendix).
With the chosen numbering of the atomic levels, the commutators
in the main equation (\ref{eq:dRdt}) remain unchanged, and only the Lindblad operator
$L' = U^{\dagger} L U$ depends on the level configuration.
Using the formulas for this operator given in the Appendix,
we again obtain that the necessary condition for the stability of
the adiabatic states $\lvert \lambda_{1} \rangle$ and $\lvert \lambda_{2} \rangle$
is $\gamma_{c}T \ll 1$.

For the $\Sigma$-system, $\gamma_{c} = \gamma_{2} + \gamma_{2\text{deph}}$
and $\gamma_{2\text{deph}}$ is the dephasing rate on this transition.
For the $V$-system,
$\gamma_{c} = \tfrac{1}{2}\left(\gamma_{1} + \gamma_{2} + \gamma_{1\text{deph}} + \gamma_{2\text{deph}}\right)$.

As for the state $\lvert \lambda_{3} \rangle$ in the $\Sigma$-system,
since the population of the second level is $\sim \Phi^{2}$,
the necessary stability condition reduces only to
$\gamma_{1}T \ll 1$.
For the $V$-system, in the case of large one-photon detunings,
the state $\lvert \lambda_{3} \rangle$ is stable, since the populations of
the upper levels $\sim \Phi^{2}$ are negligibly small. As before, the sufficient conditions for the stability of adiabatic states are the adiabaticity conditions of the interaction.

\section{Conclusion}
The necessary and sufficient conditions for the stability of adiabatic states
in different configurations of a three-level system have been investigated.
It is shown that the necessary condition for stability in all three configurations
is $\gamma_{c}T \ll 1$, where $\gamma_{c}$ is the decay rate of the coherence
induced between the levels that do not have a direct dipole transition.
The sufficient conditions for the stability of the adiabatic states are the
adiabaticity conditions of the interaction.

\begin{figure}[htb]
\includegraphics[width=0.5\textwidth]{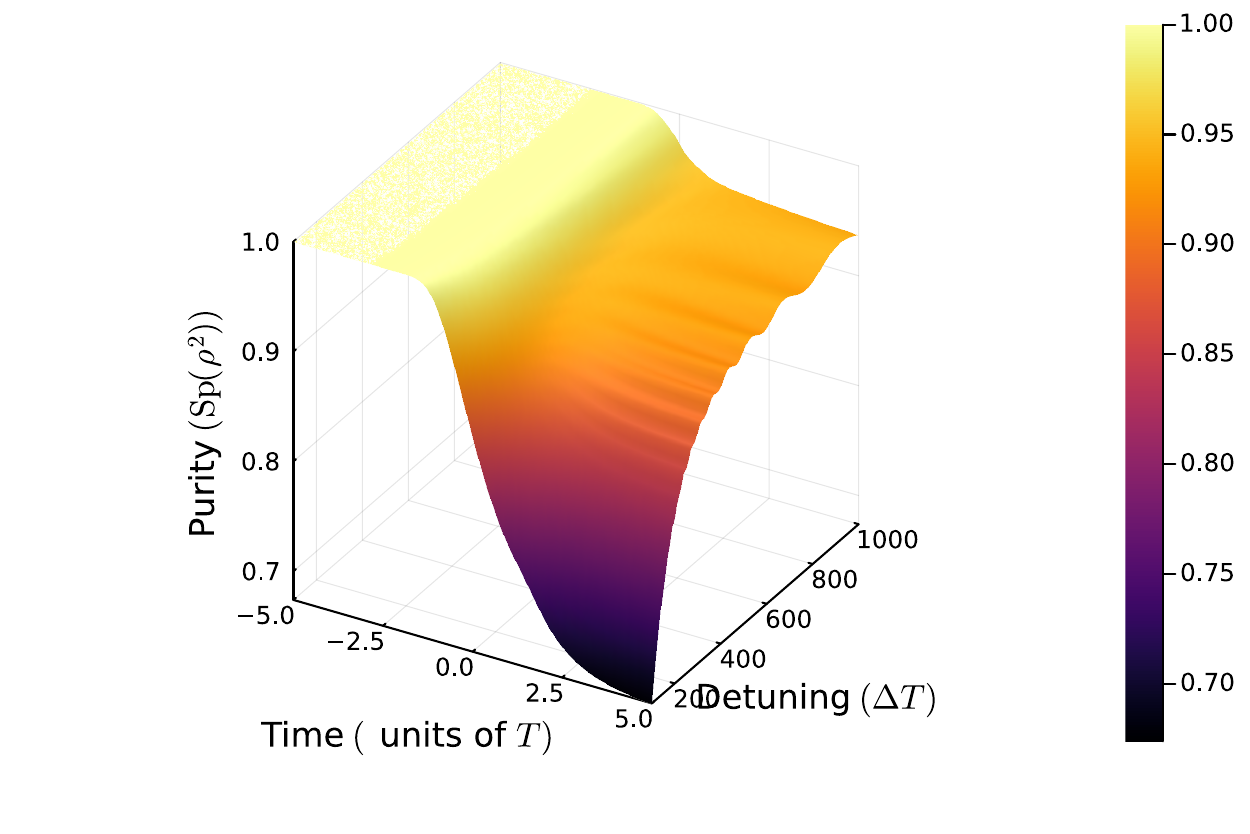}
\caption{Evolution of purity of $|\lambda_2\rangle$ with $\gamma_c$T = 0.05}
\label{Fig:sphur_delta_bstirap}
\end{figure}

Analytical estimates have been obtained in the most favorable regime of large
one-photon detunings(see Fig. \ref{Fig:sphur_delta_bstirap}). In this case, only in the $\Lambda$-system does the coherence
decay rate $\gamma_{c}$ depend on the dephasing rate of the corresponding levels,
while in other systems $\gamma_{c}$ depends on the rates of spontaneous decays.
Thus, although adiabatic states can be realized in all configurations,
in the $\Lambda$-system the stability time (interaction time) is three orders
of magnitude longer than in the $\Sigma$- and $V$-systems.

An interesting fact is that the state $\lvert \lambda_{2} \rangle$
(the so-called $b$-state \cite {28}) under large one-photon detunings is as
stable as the well-known dark state $\lvert \lambda_{1} \rangle$, see Fig.(\ref{Fig:stirap},\ref{Fig:bstirap}).
However, while in the case of the dark state $\lvert \lambda_{1} \rangle$ the intermediate level
remains unpopulated due to quantum interference, in the case of the $b$-state $\lvert \lambda_{2} \rangle$
the population of the intermediate level is almost completely and effectively eliminated owing to the large one-photon detuning.
The possibility of efficient population transfer by an intuitive sequence of pulses
(the so-called $b$-STIRAP method \cite{28,29}) is demonstrated for various values
of relaxation times, for the interaction parameters defined by the inequalities (\ref{eq:secondstb}).

As was demonstrated in \cite{28}, in a medium without dissipation the significant difference between the $b$-STIRAP method
and the traditional STIRAP is the speed of population transfer.
The $b$-STIRAP method, if relaxation processes are not being taken into account, is faster and requires less time than the STIRAP method.
This may prove to be an important advantage for the implementation of logic gates.
A combination of these two methods can also be effectively realized in multilevel systems.

Another possible application of the states $\lvert \lambda_{1} \rangle$
and $\lvert \lambda_{2} \rangle$ should be noted.
If $\theta \to \pi/4$, the state
$\lvert \lambda_{1} \rangle = \tfrac{1}{\sqrt{2}}(\lvert 1 \rangle - \lvert 2 \rangle)$,
while (to within $\Phi^{2}$ terms) the state
$\lvert \lambda_{2} \rangle = \tfrac{1}{\sqrt{2}}(\lvert 1 \rangle + \lvert 2 \rangle)$,
i.e., it represents a Hadamard gate.
For example, in alkali atoms $\gamma_{c} \sim 5\,\mathrm{kHz}$, so the lifetime
of the resulting gates will be on the order of hundreds of microseconds at room temperature.

\subsection*{Appendix}

The Lindblad operator in the general case is known to be \cite{Nathan2020}
\begin{equation}
	L = \frac{1}{2} \sum_{k \leq N^{2}-1}
	\left(2L_{k}\rho L_{k}^{\dagger} - L_{k}^{\dagger}L_{k}\rho - \rho L_{k}^{\dagger}L_{k}\right).
\end{equation}

For the $\Lambda$ system,
$L_{1} = \sqrt{\gamma_{31}}\,\sigma_{13}$,
$L_{2} = \sqrt{\gamma_{32}}\,\sigma_{23}$,
$L_{3} = \sqrt{\gamma_{3deph}}\,\sigma_{33}$,
$L_{4} = \sqrt{\gamma_{2deph}}\,\sigma_{22}$.
Here $\gamma_{31}$ and $\gamma_{32}$ are the spontaneous decay rates from the
upper level to the lower ones, with $\gamma_{31}+\gamma_{32}=\gamma_{\mathrm{sp}}$
being the total spontaneous width of the upper level.
$\gamma_{3deph}$ and $\gamma_{2deph}$ are the dephasing rates associated with
levels 3 and 2, respectively.
The transverse widths (or coherence decay rates) for the $3\!-\!1$ and $3\!-\!2$
transitions are
\begin{eqnarray}
	\Gamma_{1} = \tfrac{1}{2}(\gamma_{\mathrm{sp}}+\gamma_{3deph}),\notag \\
	\Gamma_{2} = \tfrac{1}{2}(\gamma_{\mathrm{sp}}+\gamma_{3deph}+\gamma_{2deph})\notag
\end{eqnarray}

and the decoherence rate of the lower sublevels is
$\gamma_{c} = \tfrac{1}{2}\gamma_{2deph} = \Gamma_{2}-\Gamma_{1}$.
For the $\Sigma$ system, the operators $L_{1}, L_{3}, L_{4}$ are the same
as in the $\Lambda$ system, but the operator $L_{2}$ must be replaced by the
self-adjoint $L_{2} = \sqrt{\gamma_{32}}\,\sigma_{22}$.
In this case, the coherence decay rates for the $3\!-\!2$ and $2\!-\!3$ transitions
(see Fig.~1) are
$\Gamma_{1} = \tfrac{1}{2}(\gamma_{31}+\gamma_{3deph})$,
$\Gamma_{2} = \tfrac{1}{2}(\gamma_{23}+\gamma_{31}+\gamma_{2deph}+\gamma_{3deph})$,
and the coherence decay rate is $\gamma_{21}=\gamma_{c} = \gamma_{23}+\gamma_{2deph} = \Gamma_{2}-\Gamma_{1}$.

For the $V$ system, the operators $L_{1}$ and $L_{2}$ are replaced by the
self-adjoint forms $L_{1} = \sqrt{\gamma_{13}}\,\sigma_{31}$,
$L_{2} = \sqrt{\gamma_{23}}\,\sigma_{32}$,
$L_{3} = \sqrt{\gamma_{1deph}}\,\sigma_{11}$,
$L_{4} = \sqrt{\gamma_{2deph}}\,\sigma_{22}$.
In this case, the coherence decay rates for the $1\!-\!3$ and $2\!-\!3$ transitions are
$\Gamma_{1} = \tfrac{1}{2}(\gamma_{13}+\gamma_{1deph})$,
$\Gamma_{2} = \tfrac{1}{2}(\gamma_{23}+\gamma_{2deph})$,
and the coherence decay rate is $\gamma_{21}=\gamma_{c} = \Gamma_{1}+\Gamma_{2}$.

\section{Acknowledgement}

The work was supported by the Higher Education and Science Committee of RA (project No. 1-6/IPR).
We thank the Armenian National Supercomputing Center(ANSCC) for providing the essential resources and support
that made this research possible.


\end{document}